
\input phyzzx
\tolerance=10000

{\hsize=17.5truecm \leftskip=11.5cm
{NDA-FP-3/92}\par
\vskip -4mm
{May 1992}\par}

\def\cref#1{\rlap,\attach{#1}}
\def\pref#1{\rlap.\attach{#1}}
\def\refmark#1{\attach{#1}}\def\refmark#1{\attach{#1}}
\def\odd{{\rm odd}}
\def\even{{\rm even}}
\def\ie{{\it i.e.}}

\def\eg{{\it e.g.,}}
\def\LRA{$\Longrightarrow$}
\def\LA{\ \Longrightarrow\ }
\def\e{{\rm e}}

\def\tr{{\rm tr}}
\def\co{{\cal O}}
\def\ln{{\rm ln\,}}
\def\half{{1\over 2}}
\def\nn{{N^2 \over 2}}

\title{$N=\half$ Superstring in Zero Dimension
and the Spontaneous Breakdown of the Supersymmetry}
\author{Shin'ichi Nojiri}
\address{Department of Mathematics and Physics}
\address{National Defense Academy}
\address{Hashirimizu, Yokosuka 239, JAPAN}

\abstract{We propose random matrix models which have $N=\half$
supersymmetry in zero dimension. The supersymmetry breaks down
spontaneously. It is shown that the double scaling limit can be
defined in these models and the breakdown of the supersymmetry
remains in the continuum limit. The exact non-trivial partition
functions of the string theories corresponding to these matrix models
are also obtained.}

\endpage

\REF\ri{D.J. Gross and V. Periwal\journal Phys.Rev.Lett. &60 (88)
2105, \journal Phys.Rev.Lett. &61 (88) 1517}
\REF\riii{D.J. Gross and A.A. Migdal\journal Phys.Rev.Lett. &64
(90) 127}
\REF\rvia{A. Dabholkar\journal Nucl.Phys. &B368 (92) 293}
\REF\rrrvi{G. Parisi and N. Sourlas\journal Phys.Rev.Lett.
&43 (76) 74}
\REF\rii{M.R. Douglas and S.H. Shenker\journal Nucl.Phys. &B335
(90) 635}
\REF\riv{E. Br\'ezin and V.A. Kazakov\journal Phys.Lett. &B236 (90) 144}
\REF\rv{E. Marinari and G. Parisi\journal Phys.Lett. &B240 (90) 375}
\REF\rvii{S. Nojiri\journal Phys.Lett. &B253 (91) 63 \journal
Prog.Theor.Phys. &85 (91) 671}
\REF\rvi{S. Nojiri\journal Phys.Lett. &B252 (90) 561 }
\REF\rxviii{V. Kazakov, I. Kostov and
A. Migdal\journal Phys.Lett. &B157 (85) 295}
\REF\rrrii{V. Kazakov, I. Kostov and
A. Migdal\journal Nucl.Phys. &B275 (86) 641}
\REF\rrriii{V. Boulatov, V. Kazakov, I. Kostov
and A.A. Migdal\journal Nucl.Phys. &B275 (86) 641}
\REF\rrriv{F. David\journal Phys.Lett. &B159 (85) 303}
\REF\rrrv{I. Kostov and M.L. Mehta\journal Phys.Lett. &B189 (87) 118}
\REF\rri{V.A. Kazakov\journal Nucl.Phys. &B354 (91) 614}
\REF\rria{M. Sakamoto\journal Phys.Lett. &B151 (85) 115}
\REF\rrii{S. Nojiri, preprint INS-Rep.-903, OCHA-PP-23 (1991)}
\REF\rvib{H. Nicolai\journal Phys.Lett. &B101 (80) 341\journal
Nucl.Phys. &B176 (80) 419}


\chapter{Introduction}

Superstring theory is the only one known candidate for the unified
theory of elementary particles and their interactions although there
remains many unsolved problems. One of the main problems
is to understand the mechanizm of the spontaneous breakdown of
the supersymmety. In order to solve the problem, it will be necessary
to analyze the string theory non-perturbatively. It is known that
the non-perturbative behaviors of string theories are much different
from those of usual particle theory.
The perturbative expansions of vacuum energy in critical\refmark\ri
and non-critical\refmark\riii bosonic string theories
diverge as $\sum_l l!\kappa^{2l}$, where $\kappa$ is a string
coupling constant. This implies there will appear the non-perturbative
contributions of order $\e^{{c\over \kappa}}$. Here $c$ is a
constant. Similar non-perturbative effects were also found\refmark\rvia in
Marinari and Parisi's one dimensional superstring theory\pref\rii
This situation is different from that of the usual particle
theories \eg the Yang-Mills theories
whose perturbative expansions diverge as $\sum_l (2l)!g^{2l}$
($g$ is a coupling constant) and there appear the non-perturbative
instanton effects of order $\e^{{c\over g^2}}$.
The non-perturbative effects which are characteristic of
string theories will be clarified by analyzing the \lq toy'
models of string theories.

Few years ago, a great progress was made in the
non-perturbative formulation of string theories in less than one
dimension\pref{\riii, \rii,  \riv}
These models can be regarded as an important solvable \lq toy' model,
which may give a clue to solve the dynamics of \lq realistic' string
models. In this development, string models which have a supersymmetry
in one dimension was proposed by Marinari and Parisi\pref\rv
After that, the author constructed superstring models in less than
one dimension\refmark\rvii and zero dimension\pref\rvi
The zero dimensional superstring theory is closely related to the
bosonic two dimensional gravity coupled with $c=-2$ conformal matter
($-2$ dimensional string theory)\pref{\rxviii - \rrrv}
The $-2$ dimensinal string theory can be obtained from zero
dimensional string theory by Parisi and Sourlas' dimensional
reduction mechanism\pref\rrrvi Parisi and Sourlas' mechanism
connects $D$ dimensinal theory to $D-2$ dimensional one by a nilpotent
supersymmetry, which corresponds to the supersymmetry in zero dimensional
superstring theory. The correlation functions of $D$ dimensional theory,
except vacuum amplitude, are identical with those of $D-2$ dimensional one if
the support of $D$ dimensional correlation functions is restricted to
$D-2$ hypersurface. When Parisi and Sourlas's nilpotent supersymmetry
breaks down, the $D$ dimensional theory cannot be always related with any
$D-2$ dimensional theory. Although we expect the supersymmetry in the zero
dimensional superstring can break down spontaneously, the same
non-pertubative behaviour will appear in both zero dimensional superstrings
and $-2$ dimensional bosonic strings. It has been found\refmark\rri that
the perturbative behaviour in the $-2$ dimensional string theory is similar
to that of the usual bosonic string \ie\ the perturbative expansion diverges
as $\sum_l l!\kappa^{2l}$.\foot{The author should apologize that
there was a mistake in the analysis of Ref.\rrii.}
Therefore the non-perturbative contributions of order $\e^{{c\over \kappa}}$
will also appear in zero dimensional superstring theory.

In this paper, we propose string theories which have $N=\half$
supersymmetry\refmark\rria in zero dimension. It is shown that the
supersymmetry
in these models breaks down spontaneously. The double scaling limit
can be defined in these models and the breakdown of the supersymmety
remains in the continuum limit. The partition
function of the string theories corresponding to these matrix models
can be also calculated exactly. Since these theories do not have
full Parisi and Sourlas' supersymmetry, we cannot relate these
theories  to any $-2$ dimensional theories.

\chapter{Superstring in Zero Dimension}

In this section, we briefly review the superstring theories in
zero dimension in preparation for the analysis of the $N=\half$
superstring theory.

The partition function of zero dimensional superstring theory is
given by the path integrals of an $N \times N$ hermitian matrix
$A_{ij}$ $(i,j=1,\cdots, N)$ and fermionic (anti-commuting)
$N \times N$ hermitian matrices $\overline \Psi_{ij}$, $\Psi_{ij}$
$(i,j=1,\cdots, N)$.
$$Z=(\lambda N)^{{-N^2 \over 2}}\int dAd\overline \Psi d\Psi \exp
\lambda S(A,\overline \Psi , \Psi) \ .\eqn\i$$
The action $S(A,\overline \Psi , \Psi)$ has the following form:
$$S(A,\overline \Psi , \Psi)=N \lbrace
-{1 \over 4}{\rm tr} ({\partial W(A) \over \partial A})^2
- {1 \over 2}\sum_{i,j,k,l=1}^N\overline \Psi_{ij} \Psi_{kl}
{\partial^2 W(A) \over \partial A_{ij}
\partial A_{kl}} \rbrace \ .\eqn\ii$$
Here $W(A)$ is a superpotential,
$$W(A)=\sum_{l=1}^Lg_l {\rm tr} A^l \ .\eqn\iii$$
The factor $(\lambda N)^{{-N^2\over 2}}$ in Eq.\i\ appears due to the
integration of the auxilliary field which is necessary when we
formulate the theory in superspace. The system is invariant under the
following supersymmetry transformations in zero dimension:
$$\eqalign{\delta A=&\overline \epsilon \Psi
+ \epsilon \overline \Psi \ , \cr
\delta \Psi=&
{1 \over 2}\epsilon {\partial W \over \partial A} \ , \cr
\delta \overline \Psi=&
-{1 \over 2}\overline \epsilon
{\partial W \over \partial A} \ .}\eqn\iv$$
It has been shown that the invariance under the transformation \iv\
guarantees that the free energy $F$ ($\equiv\ln Z$) of the matrix model,
{\it i.e.}, the vacuum amplitude of the corresponding string theory, vanishes
in any order of the perturbation. The supersymmetry, however, breaks down
spontaneously when the highest power of $L$ in the superpotential $W(A)$ in
Eq.\iii . This situation can be understood by using the following change of
the variables, which is called the Nicolai mapping\refmark\rvib
$$\Gamma={1 \over 2}{\partial W(A) \over \partial A} \ .\eqn\v$$
Then the partition function $Z$ can be rewritten by Gaussian integration:
$$Z=( 2\lambda N)^{{N^2 \over 2}}
\int d\Gamma \exp \lbrace - \lambda N {\rm tr} \Gamma
\Gamma \rbrace \ . \eqn\va$$
By diagonalizing the matrix $\Gamma $
(note that the matrices $A$ and $\Gamma $ can be simultaneously
diagonalized) by a unitary matrix $U$:
$$\Gamma ={1 \over \sqrt{ 2\lambda N}}U^{-1}\gamma U\ , \ \ \
\gamma={\rm diag}(\gamma_1, \gamma_2, \cdots , \gamma_N)\ , \eqn\vab$$
we obtain,
$$Z=\int \prod_{n=1}^N d\gamma_n
\prod_{m>l} (\gamma_m-\gamma_l)^2
\exp (-{1 \over 2}\sum_{k=1}^N \gamma_k^2)\ .\eqn\vb$$
When $L=\even$ Eq.\vb\ tells that the partition function $Z$ does not depends
on any coupling constant $\lambda$ or $g_l$'s in Eq. \iii and we can normalize
$Z$ to unity \ie the free energy $F$ vanishes $F=0$. On the other hand, when
$L=\odd$, the non-perturbative partition function $Z$ vanishes $Z=0$ the free
energy $F$ goes to infinity $F\rightarrow \infty$, which is a signal of the
spontaneous breakdown of the supersymmetry. The reason why $Z$ vanishes is
that the region for the integration with respect to $\gamma_n$ in Eq. \vb\
vanishes when $L=\odd$.

In Ref.\rvi , we have found that there exists a critical point by
analyzing the behaviours of the correlation functions
$<{\rm tr} A^m >$. The critical point appears in the large $N$ limit
when the Nicolai mapping in Eq.\v\ is degenerate:
$$\Gamma =-{1 \over n g^{n-1}}(g-A)^n+{g \over n} \ .\eqn\vi$$
Here $g$ is a coupling constant. In order to obtain the expressions of
$<{\rm tr} A^m >$, we calculate the correlation functions
${1 \over N}<{\rm tr} \Gamma^m >
={1 \over (2\lambda N)^{m \over 2}}<\gamma_i^m>$, $(i=1,\cdots ,N)$.
The generating function $Z(j)$ is given by
changing the action $S$ in \ii\ by adding a source term:
$S\rightarrow S+{1 \over \lambda \sqrt N}j\gamma_i$,
$$\eqalign{{Z(j) \over Z(0)}=&\exp({1 \over 2N}j^2)
\, L^1_{N-1}(-{j^2 \over N}) \cr
=&\sum_{k=0}^{N-1}\{ {1 \over (k+1)!k!}
+{1 \over N^2}{1 \over 12k!(k-2)!}+{\cal O}({1 \over N^4}) \}j^{2k}.}
\eqn\vii$$
Here $Z(0)=Z$ is the partition function without the source term and
$L^1_{N-1}$ is Laguerre's polynomial:
$L^\alpha_n(x)\equiv \sum_{m=0}^n(-1)^m{n+\alpha \choose n-m}
{x^m \over m!}$. Note that ${Z(j) \over Z(0)}$ is an even function
with respect to ${1 \over N}$.
If we expand ${Z(j) \over Z(0)}$ with respect to $j$ by
$${Z(j) \over Z(0)}=\sum_l {C_l \over l!}j^l\ , \eqn\viii$$
the correlation functions ${1 \over N}<{\rm tr} \Gamma^m >$ is given by the
coefficients $C_l$'s:
$${1 \over N}<\tr \Gamma^l>={1 \over (\sqrt 2 \lambda)^l}C_l\ . \eqn\ix$$
The coefficients $C_l$ vanish in case $l=\odd$. By using Eqs.\vii\ and \ix ,
we can calculate the correlation functions $<{\rm tr} A^m >$ when the Nicolai
mapping is degenerate \vi ,
$$\eqalign{{1 \over g^m N}&<{\rm tr} (g-A)^m >
={1 \over N}<\tr (1-{n \over g}\Gamma)^{m \over n}>\cr
&=\sum_l {{m \over n} \choose l}C_l (-\sqrt x)^l
\sim \sum_l c_l N^{-2l}(1-x)^{-3l+{3 \over 2}+{m \over n}}\ . }\eqn\x$$
Here $x$ is defined by
$$x={2^2 n^2 \over g^2 \lambda }\ .\eqn\ixa$$
Equation \x\ tells that we can obtain finite correlation functions
up to multiplicative renormalization constants to all orders in
the ${1 \over N}$ ({\it i.e.} genus) expansion if we fix
$${\bar\kappa}^{-1}= N(1-x)^{3/2} \eqn\xi$$
then by letting $x \rightarrow 1$ as $N \rightarrow \infty$.
$\bar\kappa$ can be regarded as a renormalized string
coupling constant.

When $n=\even$ in Eq.\vi\ \ie\ $L=\odd$ in Eq.\iii ,
the partition function $Z$ vanishes, therefore the
expectation value of any operator diverges or vanishes in general and
the non-perturbative theory is ill-defined.
We expect, however, the perturbative continuum limit can be obtained
by adding a higher order term to the r.h.s. in Equation \vi .
The higher order term will be irrelevant
in the continuum limit. As an example, we consider the following
Nicolai mapping:
$$\Gamma =-{(2-A)^2 \over 4\cos t +2(2-A)\sin t }
+{2 \over 1+\cos t}\ .\eqn\xi$$
When $t=0$, this mapping coincide with that in Eq.\vi\ with $n=g=2$.
Since $\gamma_n$ in Eq.\vb\ is now integrated from $-\infty$ to $\infty$,
the partition function $Z$
does not vanish and we can normalize $Z$ to unity $Z=1$.
By calculating the correlation functions
$<{\rm tr} (2-A)^m >$, we find that
$$<{\rm tr} (2-A)^m >\sim
\sum_l c_l N^{-2l}(1-\tilde x)^{-3l+{3 \over 2}+{m \over 2}}
\eqn\xii$$
Here $\tilde x$ is defined by
$$\tilde x={2(\sqrt{4\sin t} +1) \over \lambda }\ ,\eqn\xiii$$
When $t$ vanishes, $\tilde x$ smoothly reduces to $x$ in Eq.\ixa\
with $g=n=2$. Since $Z=1$ when $t\neq 0$ and the theory with $t=0$ is
expected to belong to the same university class as the theory with
$t\neq 0$, we cannot claim that the supersymmetry breaks down in
the continuum superstring theory corresponding to the theory with
$n=2$.

\chapter{$N=\half$ Superstring and the Spontaneous Breakdown of the
Supersymmetry}

In this section, we propose string theories which have $N=\half$
supersymmetry in zero dimension. It is shown that the supersymmetries
in the continuum string theory corresponding to a special class of
these models breaks down spontaneously.

We now consider the following action,
$$S(A,\overline \Psi , \Psi)=N \lbrace -{1 \over 4}{\rm tr} V(A)^2
- \sum_{i,j,k,l=1}^N \overline \Psi_{ij} \Psi_{kl}
{\partial V_{lk}(A) \over \partial A_{ij}} \rbrace\ . \eqn\xiv$$
Here $V_{ij}(A)$ $(i,j=1,\cdots, N)$ are $N \times N$ (hermitian)
matrix which is a function of $N \times N$ hermitian matrix $A$.
If there is no superpotential $W(A)$
which satisfies $V(A)={\partial W(A) \over \partial A}$, the system is
invariant under the half of supersymmetry transformations in Eq.\iv :
$$\delta A=\epsilon \overline \Psi \ , \ \delta \Psi=
{1 \over 2}\epsilon V(A) \ , \ \delta \overline \Psi=0\ .\eqn\xv$$
We call this supersymmetry as $N={1 \over 2}$ supersymmetry.
The symmetry also guarantees that the free energy $F$ of the matrix
model, {\it i.e.}, the vacuum amplitude of the corresponding string
theory, vanishes in any order of the perturbation. The Nicolai mapping
is also given by
$$\Gamma={1 \over 2}V(A)\ .\eqn\xvi$$

We propose the $N={1 \over 2}$ superstring whose action is obtained by
modifying that of $N=1$ superstring in Eq. \ii as follows;
$${\partial W(A) \over \partial A} \LA V(A)\equiv
{{\partial W(A) \over \partial A}
\over \{1+{\lambda N \over 4r^2}\tr {\partial W(A) \over \partial A}
{\partial W(A) \over \partial A}\}^\half}\ .\eqn\xvii$$
Since any superpotential $\tilde W(A)$ cannot satisfy the equation
$V(A)={\partial\widetilde W(A) \over \partial A}$, the system does not have
$N=1$ supersymmetry but $N=\half$ supersymmetry. Since the Nicolai mapping is
given by Eq.\xvi , the integration with respect to $\Gamma$ is restricted in
the finite region $\tr \Gamma^2\leq {r^2 \over \lambda N}$.
Therefore the partition function $Z$ cannot be unity
$Z\neq 1$ and the $N=\half$ supersymmetry breaks down spontaneously.
In the following, we show that the double scaling limit can be
defined in these models and the breakdown remains in the continuum
limit.

The partition function and correlation functions $<\gamma_m^l>$
are explicitly calculated in the next section,
$$\eqalign{Z&=1-\e^{-r^2}\sum_{n=0}^{{N^2 \over 2}-1}
{r^{2n} \over n!}=1-
{\Gamma({N^2 \over 2},r^2) \over \Gamma({N^2 \over 2})}\ , \cr
<\gamma_m^l>&
={C_l \over Z}\times\{
1-\e^{-r^2}\sum_{n=0}^{{N^2 +l \over 2}-1}
{r^{2n} \over n!}\}\cr
&={1 \over (\sqrt 2 \lambda)^l}{C_l \over Z}\times\{
1-{\Gamma({N^2 +l \over 2},r^2) \over \Gamma({N^2 \over 2})}\}} \ .
\eqn\xix$$
Here $\Gamma(z,p)$ is an incomplete gamma function :
$\Gamma(z,p)=\int_p^\infty \e^{-t} t^{z-1}dt$
and $C^l$ is defined in Eq.\viii .

The matrix $A$ can be given in terms of $\Gamma$. When the Nicolai's mapping
is degenerate \vi , the matrix $A$ has the following form,
$$A=g-g\{1-{{n \over g} \Gamma \over
(1-{\lambda N \over 16 r^2}\tr\Gamma^2)^\half}\}^{1 \over n}\ .
\eqn\viiia$$
This operator, however, contains divergence since the expectation
value $<{\tr \Gamma^m \over
(1-{\lambda N \over 16 r^2}\tr\Gamma^2)^{m \over 2}}>$ diverges even for
the finite $N$. Therefore we cannot discuss the double scaling limit \etc
by analyzing the correlation functions $<{\rm tr} A^m >$. Instead of $A$,
we consider the \lq\lq regularized'' operator $\tilde A$ which is defined by
$$\tilde A=g-g\{1-{n \over g} \Gamma \}^{1 \over n} \ ,
\eqn\viiib$$
and analyze the correlation functions $<{\rm tr} \tilde A^m >$.
When $r^2$ goes to infinity, $\tilde A$ coincides with $A$.

In the large $N$ limit when $r^2$ fixed, the correlation function
$<\gamma_m^l>$ is given by,
$$<\gamma_m^l>
\buildrel N\rightarrow \infty \over \longrightarrow
{1 \over (\sqrt 2 \lambda)^l}C_l\times ({2r^2 \over N^2})^l\eqn\xx$$
Therefore if we redefine $x$ in Eq.\ixa\ by $x \LA ({2r^2 \over N^2})^2x$,
the double scaling limit can be defined for the correlation functions
$<{\rm tr} \tilde A^m >$.

When ${2r^2 \over N^2}$ is fixed, the partition function in Eq.\xix\
has a singularity at ${2r^2 \over N^2}=1$:
$$\eqalign{Z
&\buildrel N\rightarrow \infty \over \longrightarrow 0\ \
({2r^2 \over N^2}<1)\cr
&\buildrel N\rightarrow \infty \over \longrightarrow 1\ \
({2r^2 \over N^2}>1)\ .}\eqn\xxii$$
If we define $\rho$ by
$${2r^2 \over N^2}\equiv 1+{\sqrt 2 \over \Lambda}\rho\ ,\eqn\xxiii$$
where $\Lambda$ is a cut-off scale which is defined by
${1 \over N}={\kappa \over \Lambda}$,
the partition function has the following form, (see the next section)
$$Z\buildrel \Lambda\rightarrow \infty \over \longrightarrow
1-\e^{-{1 \over 8}Q^2-{1 \over 2}{\rho Q \over \kappa}
-{1 \over 2}{\rho^2 \over \kappa^2}}\times{Q \e^{{1 \over 4}Q^2+1}
\Gamma ({1 \over 4}Q^2+1)\over 2\sqrt{2\pi}
({1 \over 4}Q^2+1)^{{1 \over 4}Q^2+1}}\eqn\xxiv$$
when cut-off $\Lambda$ goes to infinty, $\Lambda\rightarrow \infty$.
Here $Q$ is defined by
$$Q\equiv -{\rho \over \kappa}+\sqrt{{\rho^2 \over \kappa^2}+4} \ .
\eqn\xxv$$
The form of Eq.\xxiv\ is exact. Since $Z\neq 1$, the supersymmetry
breaks down spontaneously: Since
${d(\ln Z) \over d\rho}\propto <{\partial S(A) \over \partial (r^2)}>$,
and $<{\partial S(A) \over \partial (r^2)}>$
is given by the expectation value of the operator which is given by the
$N=\half$ supersymmetry transformation \xv\ of another operator $\co$,
the $\rho$-dependence of $Z$ means the spontaneous breakdown of the
supersymmetry.

\chapter{The Derivation of the Partition Function and Correlation Functions}

In this section, we derive the formula \xix\ and \xxiv\ for the partition
function and correlation functions $<\gamma_n^l>$.

\section{The Drivations at finite $N$}

In this subsection, we derive the formula \xix\ of partition function
and correlation functions $<\gamma_n^l>$ at finite $N$.

The partition function is given by,
$$\eqalign{Z&=( 2\lambda N)^{N^2 \over 2}
\int_{\tr \Gamma^2\leq {r^2 \over \lambda N}}
d\Gamma \exp \lbrace - \lambda N {\rm tr} \Gamma
\Gamma \rbrace \cr
&=\int_{\sum_j \gamma_j^2\leq r^2} \prod_{n=1}^N d\gamma_n
\prod_{m>l} (\gamma_m-\gamma_l)^2
\exp (-{1 \over 2}\sum_{k=1}^N \gamma_k^2)}\eqn\ai$$
Here we have diagonalized the matrix $\Gamma $ by using Eq.\vab . The
following expression of step function $\theta(x)$, ($=1$ for $x>1$ and
$=0$ for $x<0$),
$$\theta(x)=\int_{-\infty}^\infty dk{\e^{ikx} \over k-i\epsilon}\ , \eqn\aii$$
leads to the following expression of the partition function:
$$\eqalign{Z&=\int_{-\infty}^\infty dk{\e^{ir^2k} \over k-i\epsilon}
\int_{-\infty}^\infty \prod_{n=1}^N d\gamma_n
\prod_{m>l} (\gamma_m-\gamma_l)^2
\exp (-{1+ik \over 2}\sum_{k=1}^N \gamma_k^2)\cr
&=
\int_{-\infty}^\infty dk{\e^{ikx} (1+ik)^{-{N^2 \over 2}}\over
k-i\epsilon}
( 2\lambda N)^{N^2 \over 2} \times C}\eqn\aiv$$
Here $C$ is defined
$$C\equiv \int_{-\infty}^\infty \prod_{n=1}^N dq_n
\prod_{m>l} (q_m-q_l)^2
\exp (-{1 \over 2}\sum_{k=1}^N q_k^2)\sim 1\eqn\av$$
Now we have changed the variable by $q_n\equiv \sqrt{1+ik}\times\gamma_n$
and the path of the integration with respect to $q_n$ is deformed by
$Q-\sqrt{1+ik}\times\infty, \sqrt{1+ik}\times\inftyX$
\LRA $Q-\infty, \inftyX$.
By evaluating the contribution from the poles $k=i\epsilon$ and $k=i$ in
Eq.\aiv , we obtain
$$\eqalign{
Z&=1+{i^{-{N^2 \over 2}} \over ({N^2 \over 2}-1)!}
({d \over dk})^{{N^2 \over 2}-1}{\e^ikr^2 \over k}\mid_{k=i}\cr
&=1+{i^{-{N^2 \over 2}} \over ({N^2 \over 2}-1)!}
\sum_n^{{N^2 \over 2}-1}{{N^2 \over 2}-1 \choose n}
\{({d \over dk})^{{N^2 \over 2}-1-n}{1 \over k}\}
\{({d \over dk})^n\e^ikr^2\}\mid_{k=i}\cr
&=1-\e^{-r^2}\sum_{n=0}^{{N^2 \over 2}-1}
{(Nr^2)^n \over n!}\cr
&=1-{\Gamma({N^2 \over 2},r^2) \over \Gamma({N^2 \over 2})}
}\eqn\avi$$
The expression of $<\gamma_n^l>$ is also given similarly.

\section{The Partition Function at the Large $N$ Limit}

In this subsection, we derive the expression \xxiv\ for the partition
function \xix\ in the large $N$ limit \ie , the limit that cut-off $\Lambda$
goes to infinity $\Lambda\rightarrow \infty$.

By using the definition of incomplete gamma function
$\Gamma(z,p)=\int_p^\infty \e^{-t} t^{z-1}dt$, the partition function of
Eq. \avi\ can be rewritten as follows,
$$Z=1-{1 \over \Gamma (\nn)}\int_{r^2}^\infty dt \e^{-t}t^{\nn-1} \ .\eqn\bi$$
By changing the variable $t=r^2(1+\e^v)$,
we obtain the following expression:
$$Z=1-{\e^{-r^2}(r^2)^\nn \over \Gamma (\nn)}
\int_{-\infty}^\infty dv \e^{S(v)}\ .\eqn\bii$$
Here $S(v)$ has the following form,
$$S(v)\equiv -r^2\e^v+(\nn-1)\ln (1+\e^v)+v \ .\eqn\biii$$
In the following, we show that we have the finite non-trivial partition
function when cut-off $\Lambda\rightarrow \infty$ if we define $\rho$ by
Eq.\xxiii . In order to evaluate $Z$, we define $v_0$ which satisfies the
condition $S'(v_0)=0$. The explicit form of $v_0$ is given by,
$$v_0=\ln \{{-{\rho \over \kappa}+\sqrt{({\rho \over \kappa})^2
+4(1+{\rho\sqrt 2 \over \Lambda\kappa})}\over
{\sqrt 2 \Lambda \over \kappa}(1+{\rho\sqrt 2 \over \Lambda\kappa})}\}\ .
\eqn\bvii$$
Then we find that
$$\eqalign{S^{(2)}(v_0)&=S^{(3)}(v_0)+\co ({1 \over \Lambda})
=\cdots=S^{(n)}(v_0)+\co ({1 \over \Lambda})\cr
&=-{1 \over 4}Q^2-1+\co ({1 \over \Lambda})\ . }\eqn\bviii$$
Here $Q$ is defined in Eq.\xxv .
Therefore if we define $t\equiv v-v_0$, we find
$$\eqalign{S(v)&=S(v_0)+S^{(2)}(v_0)\sum_{n=2}^\infty {t^n \over n!}
+\co ({1 \over \Lambda})\cr
&=S(v_0)+S^{(2)}(v_0)(\e^t-t-1)+\co ({1 \over \Lambda})\ .}\eqn\bx$$
By changing of the variables $s\equiv \e^t$, we find
$$\eqalign{
Z&=1-{\e^{-r^2}(r^2)^\nn \over \Gamma (\nn)}\e^{S(v_0)-S^{(2)}(v_0)}
\int_0^\infty ds \e^{S^{(2)}(v_0)s}s^{S^{(2)}(v_0)
-1}\{1+\co ({1 \over \Lambda})\}\cr
&=1-{\e^{-r^2}(r^2)^\nn \over \Gamma (\nn)}\e^{S(v_0)-S^{(2)}(v_0)}
(-S^{(2)}(v_0))^{S^{(2)}(v_0)}\Gamma (-S^{(2)}(v_0))\{1
+\co ({1 \over \Lambda})\}}\eqn\bxi$$
By using Stirling's formula $\Gamma(x)=\sqrt{2\pi}x^{x-\half}\e^{-x}
\{1+\co ({1 \over x})\}$ \etc, we find the expression \xxiv\ in
the $\Lambda\rightarrow \infty$ limit.

\chapter{Summary}

The spontaneous breakdown of Marinari and Parisi's one dimensional
string was analyized in Ref.\rvia . In this paper, we have analyzed
more simple superstring models. The string models are formulated by
using matrix models which have $N=\half$ supersymmetry in zero dimension.
The supersymmetry breaks down spontaneously. The double scaling limit can be
defined in these matrix models and the breakdown of the supersymmetry remains
in the continuum limit. The exact non-trivial partition function has been
also obtained.

We have considered $N=\half$ theory for a technical reason, \ie
the partition function and the correlation functions can be calculated
explicitly. We expect that there will be a class in $N=1$ theories whose
supersymmetries break down spontaneously.

\ack{I would like to acknowledge the discussions by K. Odaka.
I am also indebted to M. Kato and A. Sugamoto for the discussions
at the early stage. Most of this work was done at INS, Tokyo University.
I am also grateful to Prof. Terazawa and other members of the theory group
for hospitality.}

\refout

\bye